\documentclass{ws-procs9x6}
\usepackage{graphicx}
\usepackage{color}
\newcommand{\tpz}{${}^{3\!}P_0$}
\newcommand{\tso}{${}^{3\!}S_1$}
\newcommand{\tdo}{${}^{3\!}D_1$}
\newcommand{\xrm}[1]{{\textstyle \mbox{\rm #1}}}

\newcommand{\One}{1\!\!1}
\newcommand{\rse}{\mathcal{R}}
\newcommand{\bes}[1]{j^{#1}_{L_{#1}}}
\newcommand{\han}[1]{h^{(1)#1}_{L_{#1}}}

\newcommand{\tmat}[2]{T_{#1#2}^{(L_{#1},L_{#2})}}
\newcommand{\sma}{\Theta_{\mbox{\scriptsize S}}}
\newcommand{\psma}{\Theta_{\mbox{\scriptsize PS}}}
\begin{document}

\title{Understanding the \lowercase{$f_0(980)$} and \lowercase{$a_0(980)$}
masses \\ as well as their widths}

\author{E.~VAN BEVEREN}

\address{Centro de F\'{\i}sica da UC \\  Departamento de
F\'{\i}sica, Universidade de Coimbra \\ P-3004-516 Coimbra, Portugal \\
E-mail: eef@uc.pt}

\author{G.~RUPP}

\address{Centro de F\'{\i}sica e Engenharia de Materiais
Avan\c{c}ados \\ Instituto Superior T\'{e}cnico, Universidade de Lisboa \\
P-1049-001 Lisbon, Portugal \\
E-mail: george@ist.utl.pt}  


\maketitle
\abstracts{
The low and approximately equal masses of the scalar mesons $f_0(980)$
and $a_0(980)$, as well as their relatively small decay widths,
are impossible to understand in terms of standard $P$-wave quark-antiquark
states. Here, these mesons are studied in a unitarised quark-meson model,
together with the other light isoscalar scalar $f_0(500)$, as
members of a complete scalar nonet below about 1~GeV. They are shown to be
dynamical states generated by a combination of quark-confinement and
strong-decay interactions, resulting in a large breaking of
$SU(3)_{\mbox{\scriptsize flavour}}$ symmetry. This is illustrated with
several pole trajectories in the complex-energy plane as a function of the
model's decay coupling constant. \\
Also, experimental evidence is presented of a still much lighter scalar
boson called $E(38)$,  which may correspond to a novel kind of mesons
predicted by V.~N.~Gribov, as an observable manifestation of a condensate
of light quarks.
}
\section{Introduction: light scalar-meson nonet}
\label{intro}
The ground-breaking work of V.~N.~Gribov on quark confinement, published
after his death in two edited papers,\cite{Gribov91} suggested the
possibility of a condensate of light quark pairs and the consequent
existence of a new kind of mesons that would manifest a very strong breaking
of $SU(3)_{\mbox{\scriptsize flavour}}$ symmetry. In those days, the early
1990s, there were indeed two scalar mesons in the tables of the Particle
Data Group whose properties seemed to rule them out as standard quark-model
mesons, viz.\ the almost mass-degenerate and relatively narrow $f_0(975)$ and
$a_0(980)$ resonances.\cite{PDG92} Thus, Gribov and co-authors
argued\cite{Gribov93} that these two scalars could be examples of such
``novel'' mesons. Their quark-antiquark configurations would then be
$(u\bar{u}+d\bar{d})/\sqrt{2}$ for the isoscalar $f_0(975)$ and
$(u\bar{u}-d\bar{d})/\sqrt{2}$ for the isovector $a_0(980)$, under the
assumption that here the $u$, $\bar{u}$ and $d$, $\bar{d}$ flavours carry
negative kinetic energy.

However, historically the status of the light scalar mesons had been even more
complicated, with the additional very broad isoscalar $\epsilon(700)$ and
isodoublet $\kappa(725)$ included in the PDG tables as early as the late
1960s (see minireview\cite{Rupp18}). The corresponding highly unusual
pattern of scalar-meson masses and widths led R.~L.~Jaffe\cite{Jaffe77} to
propose tetraquark assignments for a complete light scalar nonet, which
would e.g.\ naturally explain the approximate mass degeneracy of $f_0(975)$
and $a_0(980)$, based on an equal four-quark content. Nevertheless, already a
year earlier the PDG had dropped the entries of the light $\epsilon$ and
$\kappa$ resonances in the tables, replacing\cite{Rupp18} them by states well
above 1~GeV. This situation was maintained till 1996, with the
reintroduction of a light $\epsilon$ in the PDG tables, now called $f_0$ under
the new naming scheme and with a huge mass range of 400--1200~MeV in order to
accommodate several conflicting analyses.\cite{Rupp18} Finally, the PDG
recovered a light $\kappa$ entry in the 2004 tables, with the new name 
$K_0^\star(800)$. Very detailed data analyses in recent years gave rise to
the further renamings $f_0(600)$ (2002), $f_0(500)$ (2012), and
$K_0^\star(700)$ (2018), besides the change to $f_0(980)$ instead of $f_0(975)$
already in 1994. Together with $a_0(980)$, these scalar mesons constitute a
complete light nonet in the current PDG tables.\cite{PDG20}

In the present paper, a unitarised quark model will be revisited
that long ago predicted\cite{Beveren86} such a light scalar-meson nonet,
namely as additional and dynamical $q\bar{q}$ states owing their existence
to an interplay between the confinement and strong-decay mechanisms (see
e.g.\ Ref.\cite{Zhou20} for very recent work on this topic).
Manifest $S$-matrix unitarity and analyticity were crucial for the model's
predictions of scalar-meson masses and widths that are still today fully
compatible with PDG limits.\cite{PDG20} The original model was
developed in coordinate space, but here we shall discuss\cite{Rupp09} a
more recent, momentum-space formulation,\cite{Beveren03a} which has the
advantage of being both algebraically and analytically solvable, besides
dealing more appropriately with non-localities. Trajectories of $S$-matrix
poles will be shown to demonstrate the uncommon behaviour of these
``novel'' mesons. Nevertheless, they should be understood as normal
positive-energy phenomena resulting from the very strong coupling between
confined $P$-wave $q\bar{q}$ states and free $S$-wave two-meson states.

As an addendum, strong experimental indications of a very light new boson,
recently observed\cite{Dubna19} at the Dubna JINR and
predicted\cite{Beveren11,Beveren12a} by us almost a decade ago, will be
shown and discussed. This tentative $E(38)$ boson may indeed be one of the
novel mesons foreseen by Gribov.

The organisation of this paper is as follows. In Sec.~\ref{coord} we very 
briefly review our unitarised quark-meson model as it was originally
formulated in coordinate space. In Sec.~\ref{RSE} the corresponding
momentum-space version is recapitulated  and results of fits to the data are
presented. Section~\ref{e38} deals with the tentative $E(38)$ boson. 
In Sec.~\ref{concl} we draw some conclusions and suggest possible future work.
For a very recent general review on the crucial role of unitarity in meson
spectroscopy, see Ref.\cite{Beveren21}.

\section{Unitarised quark-meson model in coordinate space}
\label{coord}
A non-relativistic unitarised quark-meson model, developed in the late 1970s
at the University of Nijmegen, allowed a very good reproduction of the then
known vector charmonium and bottomonium spectra.\cite{Beveren80}. It was based
on harmonic-oscillator (HO) confinement with a universal frequency $\omega$
and the \tpz\ model for strong decay, the latter represented by a spherical
delta-shell to mimic the transitions between the confined heavy
$q\bar{q}$ sector and the free open-charm or open-bottom channels. However,
the model produced too small hadronic widths, despite the generated large
real mass shifts as a consequence of unitarisation.

Therefore, a generalisation of the model was formulated\cite{Beveren83}
with a smeared-out transition potential, besides the employment of relativistic
kinematics in the open two-meson channels in order allow application to
mesons in an extended mass range, in particular those with pions in their decay
modes. Thus, all two-meson decay channels consisting of combinations of
ground-state pseudoscalar and vector mesons allowed by quantum numbers were
included. The equations were solved analytically through an approximation
of the smooth decay potential by an increasing number of delta-shells until
convergence with a precision of a few MeVs was achieved. Further
phenomenological ingredients were colour hyperfine splitting from one-gluon
exchange and an energy-dependent suppression of kinematically closed two-meson
channels. Fits were then carried out to light, heavy-light, and heavy vector
as well as pseudoscalar mesons, with globally good results for the vectors and
reasonable ones for the pseudoscalars, especially considering the small number
of parameters (8 in total) and the large variety of described mesons. The
parameters were the HO frequency $\omega=190$~MeV, a dimensionless
coupling for \tpz\ decay $g^2/4\pi=7.59$, a dimensionless overall coupling
for colour splitting $g_c=5.47$, a dimensionless peak radius of the
\tpz\ decay potential $\rho_0=0.56$, and the four constituent quark masses
\begin{equation}
m_{n} = 406\;\xrm{{MeV}} ,\;
m_{s} = 508\;\xrm{{MeV}} ,\;
m_{c} = 1562\;\xrm{{MeV}} ,\;
m_{b} = 4724\;\xrm{{MeV}} \; ,
\label{qmasses}
\end{equation}
where $m_n\equiv m_u=m_d$. Note that in this model hadronic widths were
not computed perturbatively and \em a posteriori, \em but both masses and
widths followed directly from complex resonance pole positions in a manifestly
unitary multichannel $S$-matrix. Thus, even overlapping broad resonances would
still automatically satisfy unitarity.

These features made the model ideal to study the very controversial light
scalar mesons, in view of the very unusual pattern of scalar masses and widths
and even their questionable existence as true quark-model resonances. Thus, in
Ref.\cite{Beveren86} a complete light scalar-meson nonet was found, as
additional and dynamical resonances calculated in the above multichannel
model\cite{Beveren83} with completely unaltered parameters, Now, 34 years
later, those predictions for pole positions (in MeV), viz.\
$\epsilon(470-i208)$ ($f_0(500)$), $S^\star(994-i20)$ ($f_0(980)$),
$\kappa(727-i263)$ ($K_0^\star(700)$), and $\delta(968-i28)$ ($a_0(980)$) are
still fully within the latest PDG\cite{PDG20} limits. In Fig.~1
we also show the computed
\begin{figure}[!b]
\label{pipi}
\begin{center}
\includegraphics[trim = 50mm 175mm 50mm 45mm,clip,width=11.3cm,angle=0]
{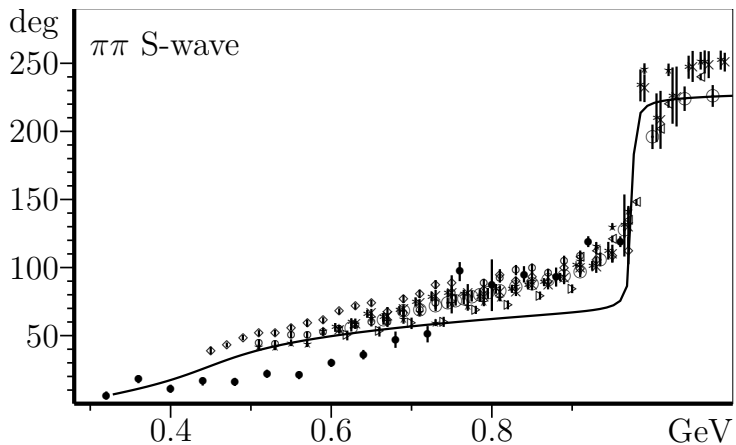}
\end{center}
\caption{$S$-wave $\pi\pi$ phases predicted in
\protect Ref.\protect\cite{Beveren86} (see this reference for data).}
\end{figure}
$S$-wave $\pi\pi$ phase shift, obtained without any fit, together with the
then available data (see Ref.\cite{Beveren86}). Besides the mentioned light
scalar mesons, the model also predicted another complete light scalar
nonet at roughly 1.3--1.5~GeV, corresponding to a few then already
observed\cite{PDG86} $0^{+(+)}$ resonances and also in agreement with
expectations from the static quark model for standard $P$-wave $q\bar{q}$
states.

\section{Scalar mesons in a unitarised momentum-space model}
\label{RSE}
The coordinate-space models presented in the previous section had achieved
a remarkable good description of the whole meson spectrum, in particular
the light scalar mesons. It allowed for the first time to reproduce these
controversial resonances as a complete scalar nonet in the context of the quark
model, yet not as regular quark-antiquark mesons but instead as dynamical
complex-energy poles that cannot be linked in a simple way to the discrete
confinement spectrum. However, especially the more general model of
Ref.\cite{Beveren83} was based on a very complicated derivation of the
multichannel $S$-matrix and so not very accessible to experimentalists in
their quest for unitary multichannel tools to analyse the
spectroscopic data.

Thus, we first developed\cite{Beveren01} a simple yet non-perturbative
momentum-space version of the model, which in the limit of a small decay
constant gives rise to Breit--Wigner-like expressions for resonances. A
detailed derivation of this so-called Resonance-Spectrum Expansion (RSE),
employing like in Ref.\cite{Beveren80} a single spherical delta-shell in
coordinate space to mimic the \tpz\ decay potential, can be found in
Ref.\cite{Beveren03a}. The RSE model, for one confined quark-antiquark
channel coupled to one free meson-meson (MM) channel, was applied in
Ref.\cite{Beveren01} to show that the then still unlisted\cite{PDG00} scalar
$\kappa$ resonance could be easily extracted from the available experimental
data on $S$-wave $K\pi$ scattering, with a pole position that is compatible
with the present PDG limits for the $K_0^\star$(700) resonance.\cite{PDG20}
Exactly the same model was the first successful description\cite{Beveren03b}
of the then newly discovered and very enigmatic scalar charm-strange meson
$D_{s0}^\star$(2317),\cite{PDG20} the only change being the replacement of a
light-quark mass by a charm-quark mass.

Now, this simple model was adequate to describe $K_0^\star(700)$ and
$D_{s0}^\star$(2317), because both resonances couple to only one 
quark-antiquark state and all but the lowest MM channel that couple
lie at much higher energies. However, in general there are more relevant
decay modes and in many cases also more than one quark-antiquark channel,
e.g.\ in the case of vector mesons, where there are both \tso\ and \tdo\
$q\bar{q}$ components. The first generalisation was presented and used
in Ref.\cite{Beveren06}, namely to model the light scalar mesons with 
all pseudoscalar-pseudoscalar (PP) decay channels included and also
both $n\bar{n}\equiv(u\bar{u}+d\bar{d})/\sqrt{2}$ and $s\bar{s}$ 
quark-antiquark channels in the isosinglet case. This multichannel model
is also manifestly unitary, but only when the dependence of the channel
couplings on the radial quantum number $n$ (see below) is the same for all
considered MM channels, which is the case for $S$-wave PP channels.
A further generalisation\cite{Rupp09} was therefore due in order to cover
all possible $n$ dependencies, which is the model employed in the present
study.

The momentum-space model allows a very simple and instructive diagrammatic 
representation of the $T$-matrix for MM scattering with non-exotic
quantum numbers, viz.\cite{Rupp09} \\[2mm]
\begin{tabular}{ccccc}
\raisebox{5mm}{$T\;\;\;=$} &
\includegraphics[trim = 15mm 0mm 0mm 0mm,clip,height=1.2cm,angle=0]
{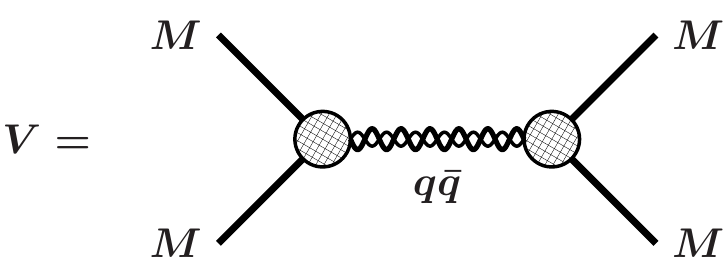} &
\raisebox{5mm}{$+$} &
\includegraphics[trim = 24mm 0mm 0mm 0mm,clip,height=1.2cm,angle=0]
{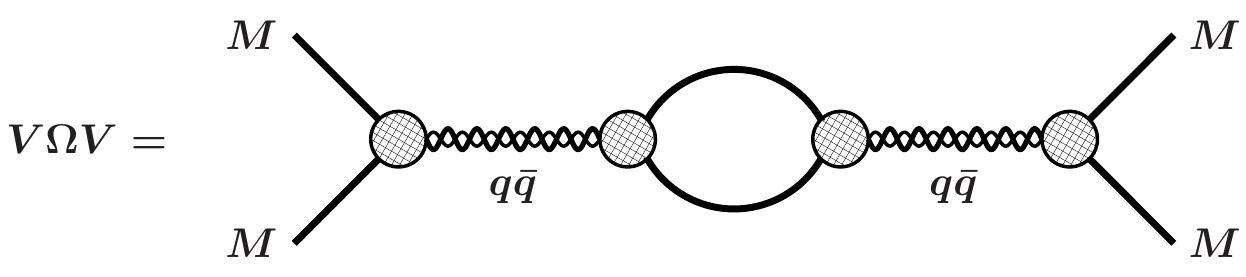} &
\raisebox{5mm}{$+\;\;\;\ldots$ \,.} \\[2mm]
\end{tabular}
Here, the first diagam on the right-hand side stands for the effective MM
potential $V$ generated by the MM$q\bar{q}$ vertices and the $q\bar{q}$
propagator in the intermediate state. The wiggly lines indicate that this
is not just one $q\bar{q}$ state but a whole, in principle infinite tower
of $q\bar{q}$ states with the same quantum numbers, and therefore a kind of
Regge propagator.\cite{Beveren09} The second diagram is the once-iterated
$V$, with an MM loop in between, and so the second term in a Born series.
Since $V$ is separable, as it is only based on $s$-channel exchanges, the
whole series can be summed up easily, giving rise to a closed-form expression
for the $T$-matrix. The explicit momentum-space formulae for $V$ 
read\cite{Coito09}
\begin{equation}
V_{ij}^{L_i,L_j}(p_i,p'_j;E)=
\lambda^2j^i_{L_i}(p_ia)\,\mathcal{R}_{ij}(E)\,j^j_{L_j}(p'_ja) \;,
\label{inter}
\end{equation}
\begin{equation}
\mathcal{R}_{ij}(E)=\sum_{i_{q\bar{q}}=1}^{N_{q\bar{q}}}\sum_{n=0}^{\infty}
\frac{g^i_{(i_{q\bar{q}},n)}g^j_{(i_{q\bar{q}},n)}}{E-E_n^{(i_{q\bar{q}})}}\;,
\label{rse}
\end{equation}
where the RSE propagator $\mathcal{R}$ contains an infinite tower of
$s$-channel bare $q\bar{q}$ states, corresponding to the discrete spectrum
of an arbitrary confining potential. Here, $E_n^{(i_{q\bar{q}})}$ is the energy
level of the $n$-th recurrence in the $i_{q\bar{q}}$-th $q\bar{q}$ channel,
with $N_{q\bar{q}}$ the number of $q\bar{q}$ channels having the same quantum
numbers, and $g^i_{(i_{q\bar{q}},n)}$ is the corresponding coupling to the
$i$-th MM channel. Furthermore, in Eq.~(\ref{inter}), $\lambda$ is the overall
coupling constant for \tpz\ decay, and $\bes{i}(p_i)$ and $p_i$ are the
$L_i$-th order spherical Bessel function and the (relativistically defined)
off-shell relative momentum in MM channel $i$, respectively. The spherical
Bessel function originates in our string-breaking picture of OZI-allowed decay,
being just the Fourier transform of a spherical delta-shell of radius $a$.
The channel couplings $g^i_{(i_{q\bar{q}},n)}$ in Eq.~(\ref{rse}) are computed
following the formalism developed in Ref.\cite{Beveren82}, namely from
overlaps of HO wave functions for the original $q\bar{q}$ pair, the created
\tpz\ pair, and the quark-antiquark states corresponding to the outgoing
two mesons. In most cases, this method produces the same couplings for
ground-state mesons as the usual point-particle approaches, but also
provides a clear prescription for excited mesons, with the additional
advantage of always resulting in a finite number of non-vanishing couplings.
Because of their fast decrease for increasing radial quantum number $n$,
practical convergence of the infinite sum in Eq.~(\ref{rse}) is achieved by
truncating it after at most 20 terms.

With this effective energy-dependent MM potential, the $T$-matrix reads
explicitly\cite{Coito09}
\begin{eqnarray}
\lefteqn{\tmat{i}{j}(p_i,p'_j;E)=-2a\lambda^2\sqrt{\mu_ip_i}\,\bes{i}(p_ia)
\times} \nonumber \\
&&\hspace*{-1pt}\sum_{m=1}^{N}\rse_{im}\left\{[\One-\Omega\,\mathcal{R}]^{-1}
\right\}_{\!mj}\bes{j}(p'_ja)\,\sqrt{\mu_jp'_j} \; ,
\label{tmat}
\end{eqnarray}
with the loop function
\begin{equation}
\Omega_{ij}(k_j)=-2ia\lambda^2\mu_jk_j\,\bes{j}(k_ja)\,\han{j}(k_ja)\,
\delta_{ij}\;, 
\label{omega}
\end{equation}
where $\han{j}(k_ja)$ is the spherical Hankel function of the first kind,
$k_j$ and $\mu_j$ are the on-shell relative momentum and reduced mass in
MM channel $j$, respectively, and the matrix $\mathcal{R}(E)$ is
given by Eq.~(\ref{rse}). Note that no regularisation is needed in this
all-orders model, since the Bessel functions at the vertices make the meson
loops finite.

Now we are in a position to apply this model to scalar mesons just like in 
Ref.\cite{Rupp09}. However, since here we focus on the scalars $f_0(980)$ and
$a_0(980)$, we do not deal with $K_0^\star(700)$ and only consider the
mixed isoscalars $f_0(500)$/$f_0(980)$ and the isovector $a_0(980)$. The
following PP, vector-vector (VV), and scalar-scalar (SS) two-meson decay
channels are included: \\[1mm]
{\boldmath{$f_0(500)/f_0(980)$}:} \\
\hspace*{\fill}\parbox[t]{11cm}{
$\pi\pi$, $K\bar{K}$, $\eta\eta$, $\eta\eta^\prime$, $\eta^\prime\eta^\prime$,
$\rho\rho$, $\omega\omega$, $K^\star\bar{K}^\star$, $\phi\phi$,
$(\rho\rho)_{L=2}$, $(\omega\omega)_{L=2}$, $(K^\star\bar{K}^\star)_{L=2}$,
$(\phi\phi)_{L=2}$, $f_0(500)f_0(500)$, $f_0(980)f_0(980)$,
$K_0^\star(700)\bar{K}_0^\star(700)$, $a_0(980)a_0(980)$} \\[2mm]
{\boldmath{$a_0(980)$}:} \\
\hspace*{\fill}\parbox[t]{11cm}{
$\pi\eta$, $K\bar{K}$, $\pi\eta^\prime$, $\rho\omega$, $K^\star\bar{K}^\star$,
$(\rho\omega)_{L=2}$, $(K^\star\bar{K}^\star)_{L=2}$, $a_0(980)f_0(500)$,
$K_0^\star(700)\bar{K}_0^\star(700)$} \\[2mm]
Note that all channels are in $S$-waves, except for those marked with
$L\!=\!2$. It may also be considered strange that some decay channels
are included with the same mesons as the ones that we are going to
describe, which looks like a (partial) bootstrap. However, by including
these channels we try to mimic e.g.\ the observed\cite{PDG20} decay modes
$f_0(1370)\to2(\pi\pi)_{\mbox{\scriptsize $S$-wave}}$ and 
$a_0(1450)\to a_0(980)\pi\pi$, in which the two-pion final states are
probably dominated by the $f_0(500)$ resonance. As for the confined sector,
we only need the $(u\bar{u}-d\bar{d})/\sqrt{2}$ channel in the isovector case,
However, for isoscalars $n\bar{n}=(u\bar{u}+d\bar{d})/\sqrt{2}$ and
$s\bar{s}$ can mix and couple to both $f_0(500)$ and $f_0(980)$. Part of this
mixing will inevitably proceed via $K\bar{K}$ loops, but some mixing purely
on the quark level should also be considered. A strong indication
comes from the relatively large hadronic decay width of the scalar charmonium
state $\chi_{c0}(1P)$, viz.\ about 10~MeV,\cite{PDG20} in spite of being due
to OZI-forbidden processes only. For comparison, the total width of the vector
charmonium state $J/\psi$ is only 0.093~MeV. This difference is usually
attributed to $C$-parity, which allows hadronic decays of a $J^{PC}=0^{++}$
state via $q\bar{q}$ annihilation to two intermediate gluons, whereas for a
$J^{PC}=1^{--}$ state three gluons are needed to conserve $C$-parity.
Thus, in our coupled $f_0(500)$/$f_0(980)$ case, we take two $q\bar{q}$
channels that are already mixtures of pure $n\bar{n}$ and $s\bar{s}$ states, 
with bare ground-state masses $M_1$ and $M_2$ given by the quadratic mass
formulae\cite{Scadron13}
\begin{eqnarray}
M^2_{n\bar{n}} & = & (M_1\cos\sma)^2 \; + \; (M_2\sin\sma)^2 \; , 
\label{m1} \\
M^2_{s\bar{s}}\: & = & (M_1\sin\sma)^2 \; + \; (M_2\cos\sma)^2 \; ,
\label{m2}
\end{eqnarray}
where $\sma$ is the scalar mixing angle, to be fitted next. In the
isovector case, one $(u\bar{u}-d\bar{d})/\sqrt{2}$ channel is naturally
sufficient. In both cases, we take the bare RSE energy levels to be given by
an HO spectrum with constant frequency $\omega$, like e.g.\ in
Refs.\cite{Beveren06,Rupp09}, i.e.,
\begin{equation}
E_{\mbox{\scriptsize HO}} \; = \; (2n+\ell+3/2)\,\omega+m_{q_1}+m_{q_2} \; ,
\label{ho}
\end{equation}
with $n$ and $\ell$ the radial and orbital quantum numbers, respectively,
and $m_{q_1}$, $m_{q_2}$ the constituent quark masses. So $m_{q_1}+m_{q_2}$
is taken as $M_1$ from Eqs.~(\ref{m1},\ref{m2}) for $f_0(500)$ and its higher
recurrences, as $M_2$ for $f_0(980)$, and as
$M_{n\bar{n}}=2m_n$ for $a_0(980)$. A final phenomenological ingredient of
our present model is an additional subthreshold suppression of kinematically
closed MM channels, as the loop function in Eq.~(\ref{omega}) turns out to be
insufficient for that purpose in the fits to the data. Such a suppression was
dealt with in previous work by multiplying the overall coupling $\lambda$ with
a factor $E/T_i$ (see e.g.\ Ref.\cite{Beveren83}), with $T_i$ the threshold
mass of the relevant channel,\cite{Beveren83} or by a form factor
$\exp(-\alpha|k_i|^2)$ (see e.g.\ Ref.\cite{Beveren06}), where $k_i$ is
the channel's relative momentum and $\alpha$ an adjustable constant. Here, we
opt for the latter solution but written analytically as $\exp(\alpha k_i^2)$,
which allows a more rigorous continuation into the complex-energy plane when
searching for poles (see below). This damping constant is taken as
$\alpha=6$~GeV$^{-2}$ in the isoscalar case and $\alpha=3$~GeV$^{-2}$ for the
isovector.

The few free parameters of the model are fitted to $S$-wave $\pi\pi$ phase
shifts from threshold up to 1.6~GeV in the isoscalar case and to the $a_0(980)$
line shape for the isovector (see Ref.\cite{Beveren06} for the used data).
For stability reasons, we do not fit $\alpha$, but select its optimum value
by hand in either case. Note that we take the values of $m_n$, $m_s$, and
$\omega$ unchanged as in all our previous work, with $\omega=190$~MeV and
$m_n$, $m_s$ given in Eq.~(\ref{qmasses}). Thus, the isoscalar fit parameters
are the coupling constant $\lambda$, the decay radius $a$, and the scalar and
pseudoscalar mixing angles $\sma$ and $\psma$, the latter being needed to 
properly determine the couplings to channels involving the $\eta$ and/or
$\eta^\prime$ mesons. In the isovector case, the free parameters are
$\lambda$, $\psma$, $a_{n\bar{n}}$ and $a_{s\bar{s}}$,  where the
latter two decay radii are taken different for $n\bar{n}$ and $s\bar{s}$
pair creation. For only one $q\bar{q}$ channel, this is possible without
changing the structure of Eqs.~(\ref{inter}--\ref{omega}), which is not 
feasible when there are more than one $q\bar{q}$ channels. The fitted
parameters are: \\[1mm]
{\boldmath{$f_0(500)/f_0(980)$}:} \\[0.5mm]
$\lambda=3.557$ GeV$^{-1/2}$, $a=3.417$ GeV$^{-1}$, $\sma=17.54^\circ$,
$\psma=41.78^\circ$ \\[1mm]
{\boldmath{$a_0(980)$}:} \\[0.5mm]
$\lambda=2.132$ GeV$^{-1/2}$, $a_{n\bar{n}}=3.944$ GeV$^{-1}$,
$a_{s\bar{s}}=2.900$ GeV$^{-1}$, $\psma=42.33^\circ$ \\[1mm]
First of all, it is very reassuring that in both fits $\psma$ comes out
close to $42^\circ$, which is the value favoured in many experimental
and theoretical analyses.\cite{Scadron13} Moreover, also $\sma$ is near
what one would expect from estimates\cite{Scadron13} based on quadratic
mass formulae and certain decay rates, namely roughly $20^\circ$. Finally,
$a$ comes out right in the middle of $a_{n\bar{n}}$ and $a_{s\bar{s}}$,
which lends additional support to the physical significance of the fits.
The fact that $\lambda$ comes out considerably larger in the isoscalar
case than for the isovector may due be to a deficient handling of the
important $4\pi$ modes\cite{PDG20} above roughly 1~GeV. This is also
indicated by the fitted phase shifts, which reasonably reproduce the
experimental ones up to about the $K\bar{K}$ threshold, but thereabove
first show an undershoot followed by an overshoot. We shall come back
to this point in Sec.~\ref{concl}. As for the fitted line shape of
$a_0(980)$, we get deviations of less than 2\% from 900~MeV all the way
up to near the $\pi\eta^\prime$ threshold at 1093~MeV. It lies outside
the scope of the present paper to show details of the fits, since the
main issue here is the nature of the $f_0(980)$ and $a_0(980)$ resonances.
For the fitted parameters, we find the following most important poles
(in MeV): \\[1mm]
{\boldmath{$f_0(455-i232)$}, $f_0(1007-i17.4)$ (second sheet),
$f_0(1290-i131)$ \\
$a_0(1017-i39.6)$ (second sheet), $a_0(1341-i285)$} \\[1mm] 
Poles on the second Riemann sheet are defined here as quasi-bound states
with respect to the $K\bar{K}$ threshold, i.e., with positive imaginary
part in the corresponding relative momentum. All these scalar-meson pole
positions are very reasonable, especially for the lowest states.\cite{PDG20}

In order to better understand the nature of these resonances, we now
show how they move in the complex-energy plane as a function of the 
overall coupling $\lambda$, as displayed in Figs.~2--4. Note that in 
Figs.~3 and 4 solid lines represent trajectories on what we call
``natural'' Riemann sheets, where the imaginary part of the relative
momentum with respect to each threshold is positive or negative, according as
the pole lies below or above that threshold, respectively. All other
(parts of) trajectories are drawn as dashed lines in those figures.
In Fig.~2 the trajectory of
\begin{figure}[!t]
\label{sigma}
\begin{tabular}{c}
\hspace*{-4mm}
\includegraphics[trim = 0mm 4mm 0mm 0mm,clip,width=11.5cm,angle=0]
{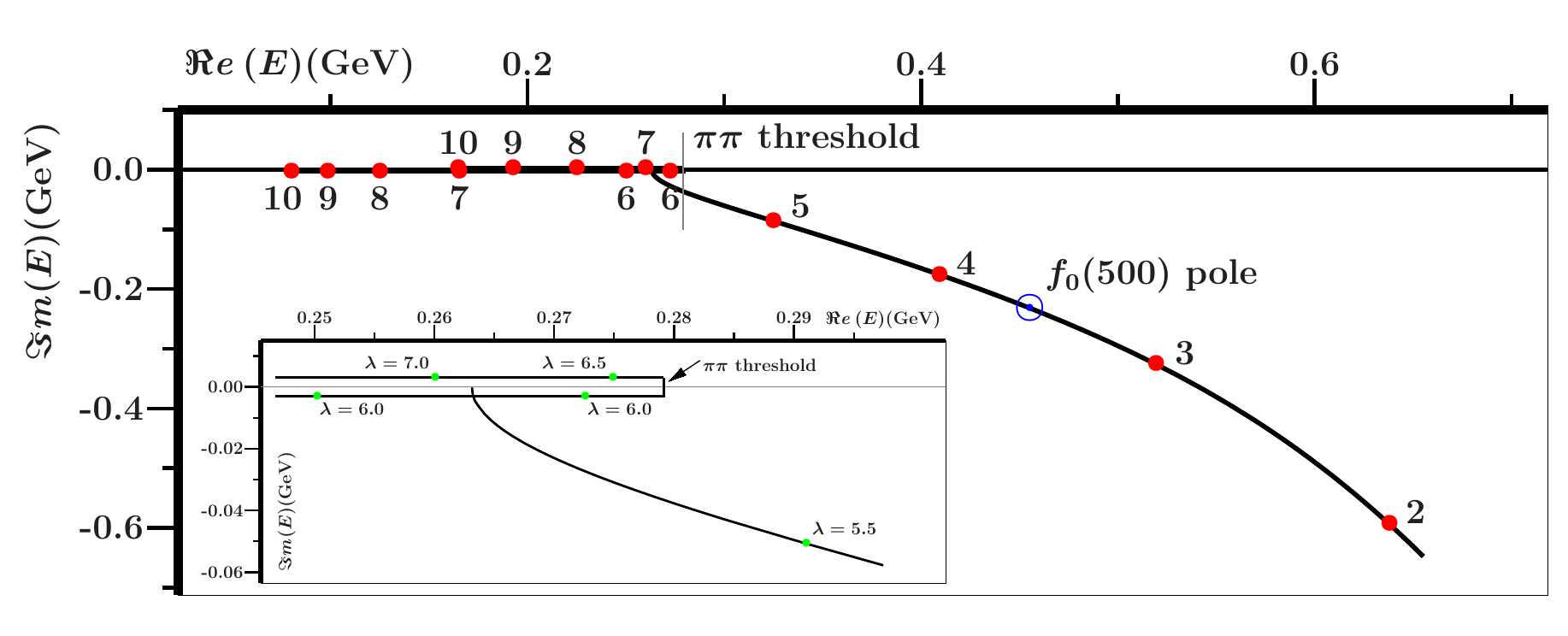} \\
\hspace*{1mm}
\includegraphics[trim = 0mm 0mm 0mm 10mm,clip,width=10cm,angle=0]
{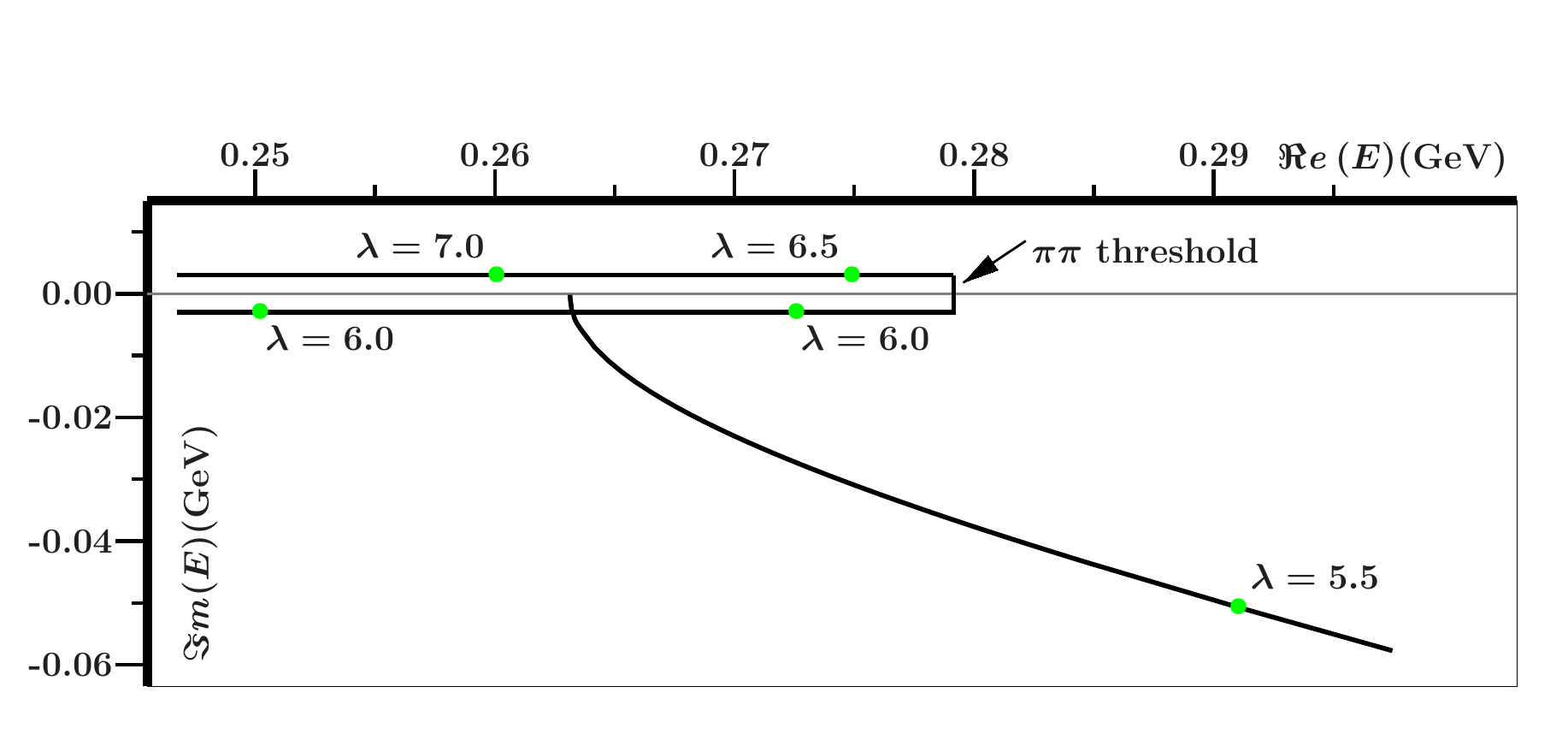} 
\end{tabular}
\caption{$f_0(500)$ pole trajectory as a function of $\lambda$. The open
circle corresponds to the fitted $\lambda$ value. The inset (enlarged
underneath) shows details of (virtual) bound states, for clarity depicted
slightly (below) above the real axis. Also see text.}
\end{figure}
the $f_0(500)$ pole is depicted, with the inset displaying its behaviour
below the $\pi\pi$ threshold. We see the pole remaining complex below
threshold, as it should\cite{Taylor72} for $S$-waves, and then turning into
two virtual bound states, one of which moves upwards again and then becomes
a true bound state. For clarity, in the inset the virtual bound states are 
shown somewhat below the real axis and the bound state above the axis.
In the limit of small coupling, the pole moves deep into the complex
plane, as a manifestation of its dynamical and not intrinsic nature. Next we
inspect the $f_0(980)$ and $f_0(1370)$ trajectories in Fig.~\ref{f0s}.
\begin{figure}[!t]
\label{f0s}
\hspace*{2mm}
\includegraphics[trim = 20mm 134mm 10mm 30mm,clip,width=11.5cm,angle=0]
{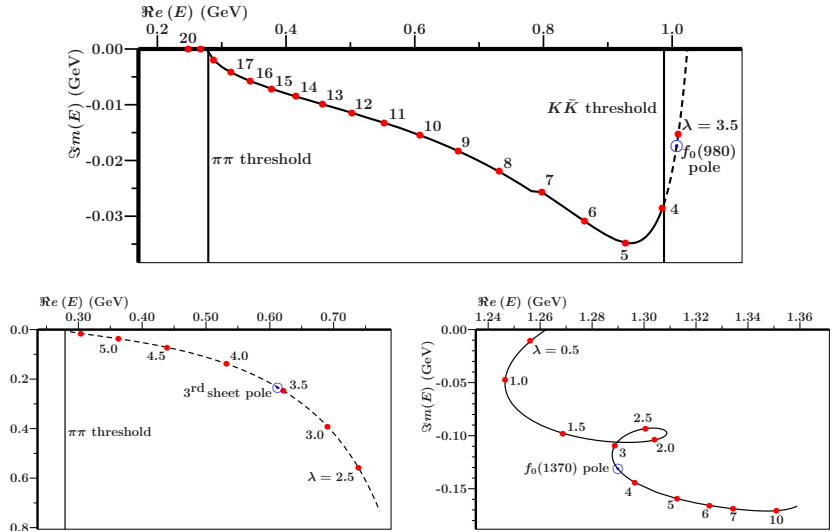}
\caption{$f_0(980)$ pole trajectories as a function of $\lambda$ on
second sheet (upper) and third sheet (lower left); $f_0(1370)$
trajectory (lower right). Also see text.}
\end{figure}
\begin{figure}[!t]
\label{a0s}
\begin{tabular}{cc}
\hspace*{-3mm}
\includegraphics[trim = 0mm 0mm 0mm 0mm,clip,width=5.5cm,angle=0]
{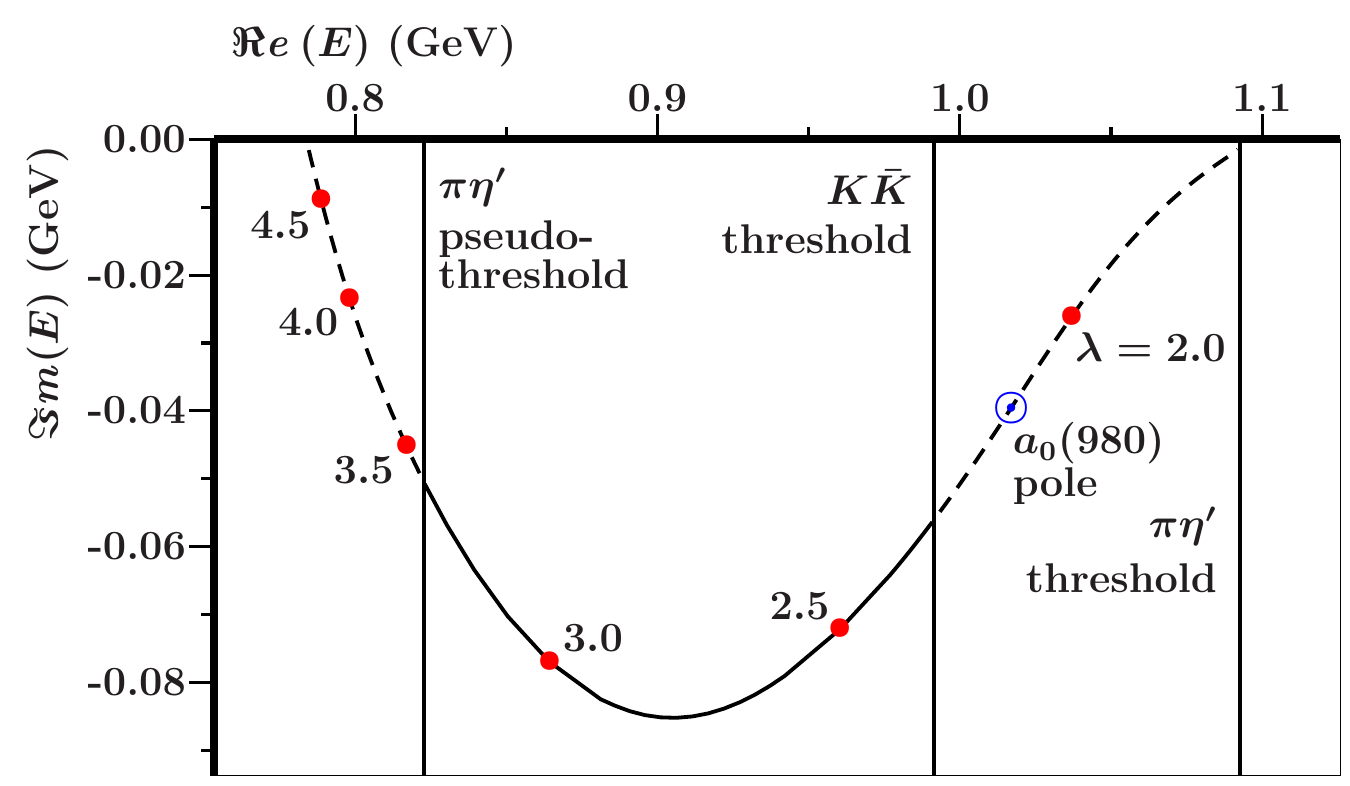} &
\hspace*{-3mm}
\includegraphics[trim = 0mm 0mm 0mm 0mm,clip,width=5.5cm,angle=0]
{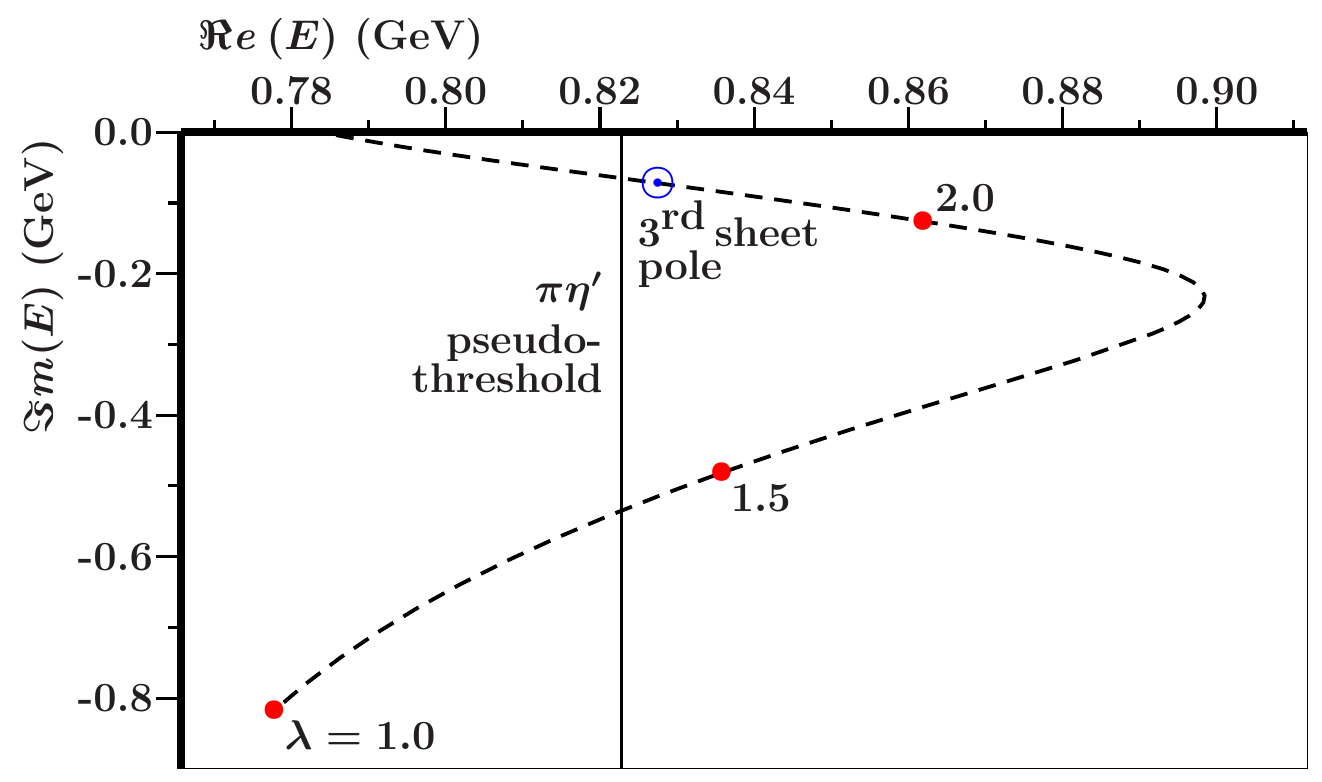} \\
\hspace*{-1mm}
\includegraphics[trim = 0mm 0mm 0mm 0mm,clip,width=5.5cm,angle=0]
{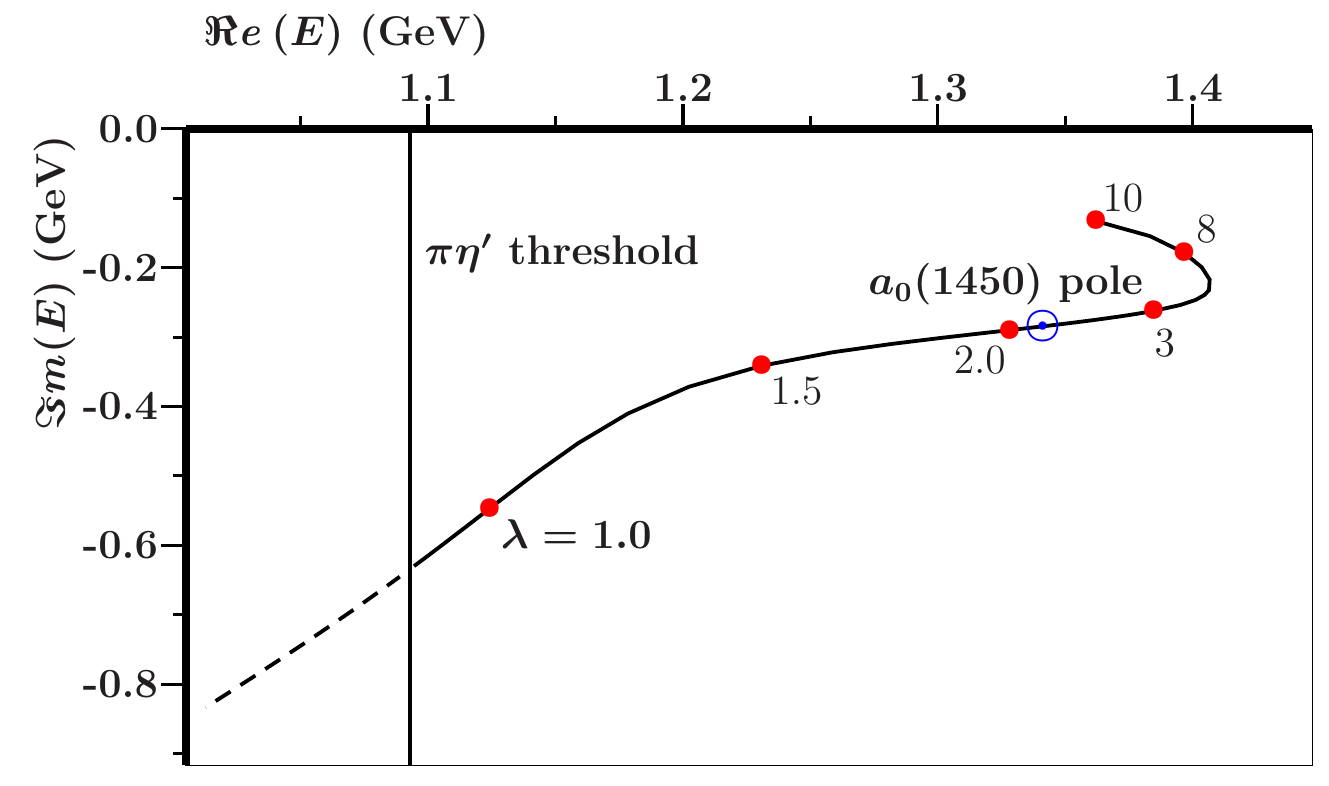} &
\hspace*{-4mm}
\includegraphics[trim = 0mm 0mm 0mm 0mm,clip,width=5.5cm,angle=0]
{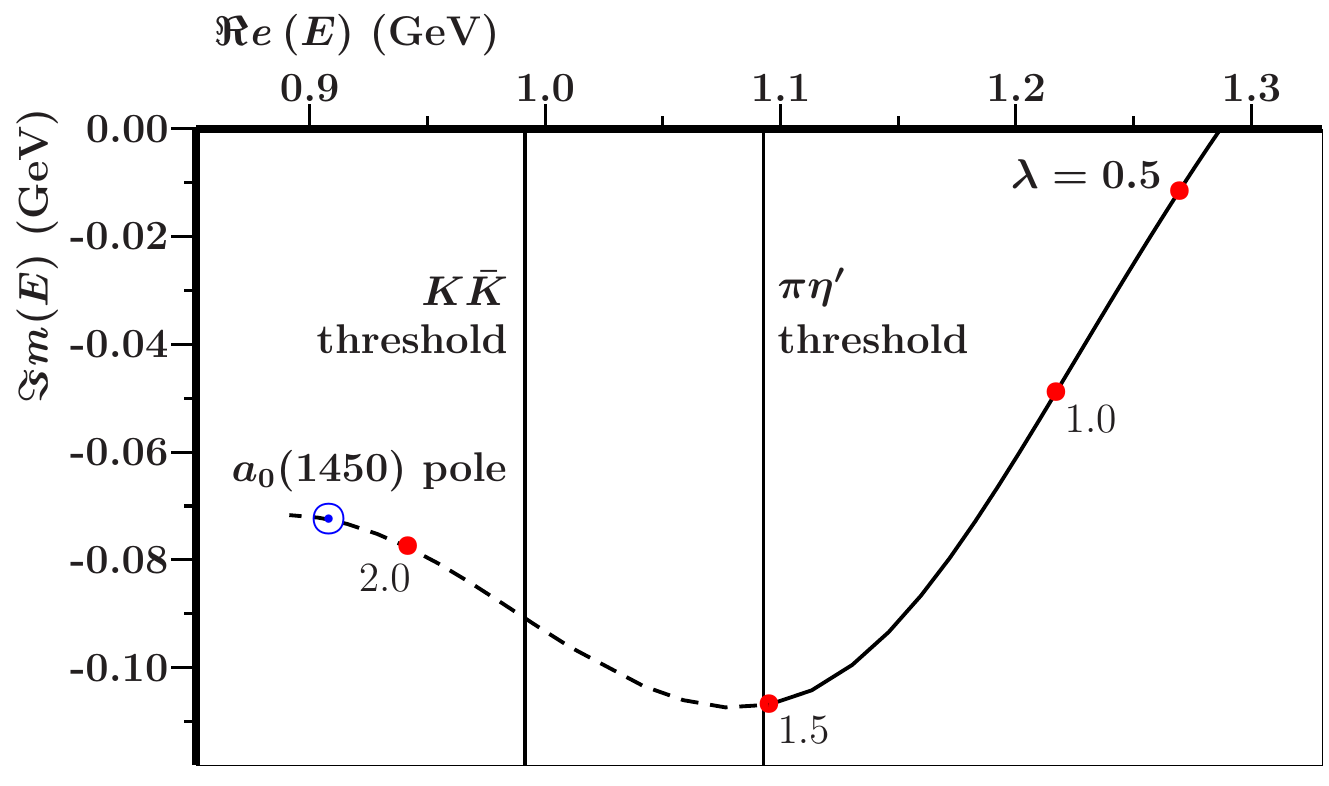} 
\end{tabular}
\caption{$a_0(980)$ pole trajectories as a function of $\lambda$ on
second sheet (upper left) and third sheet (upper right); $a_0(1450)$
pole trajectories: dynamical state (lower left) and intrinsic state
(lower right). Also see text.}
\end{figure}
Note that the pole representing $f_0(980)$ ends up as a second-sheet pole above
the $K\bar{K}$ threshold, but is still very close to the physical region.
Also, this resonance cannot be directly linked to any intrinsic $q\bar{q}$
state either, nor can the corresponding third-sheet pole, so both must
be considered dynamical. Only the $f_0(1370)$ pole is clearly intrinsic,
though with a highly non-perturbative and non-linear behaviour.
The final set of trajectories concern the isovectors, depicted in
Fig.~4. The dominant $a_0(980)$ pole lies again on the second sheet
and above the $K\bar{K}$ threshold. However, this time there is also a
third-sheet pole not very far from the physical region. Both $a_0(980)$ poles
are clearly dynamical. The surprise here is that the $a_0(1450)$ resonance
also shows up as a dynamical pole, though with a too large imaginary
part.\cite{PDG20} On the other hand, the pole of the bare $a_0(1450)$
state moves very far downwards into a non-physical region and so will
have hardly any influence on the amplitudes for the fitted value of $\lambda$.

Summarising our findings above, it should be clear by now that the light
scalar mesons are different from the other mesons in the sense that no
direct connection to quark-antiquark states in the discrete confinement
spectrum is possible. Their dynamical nature as $q\bar{q}$ states with large
MM components has also been confirmed\cite{Dudek} in recent years through
fully unquenched lattice calculations employing both $q\bar{q}$ and 
MM interpolators. In the present model, also the $a_0(1450)$ resonance
turns out to be of a dynamical nature. However, some caution is due as
no scattering data in that energy region are available and so we were
forced to fit the model parameters only with the $a_0(980)$ line shape.
Moreover, for some resonances to be considered dynamical or intrinsic
may hinge upon model specifics or minor changes of parameters, as
e.g.\ in the case of $\chi_{c1}(3872)$.\cite{PDG20,Coito11,Coito13}

\section{Evidence of a very light boson with a mass of 38 MeV}
\label{e38}
The light scalar mesons described above manifest a very strong breaking of
$SU(3)_{\mbox{\scriptsize flavour}}$ symmetry, in particular $f_0(980)$
and $a_0(980)$, which are almost degenerate in mass yet have very different
(dominant) quark contents. However, they do not appear to correspond to the
``novel'' mesons with negative kinetic energy as proposed\cite{Gribov93} by
Gribov and collaborators. This leaves open the question whether observable
mesonic states from a condensate of light quarks may exist. For that matter,
in Ref.\cite{Beveren11} we proposed the existence of a very light scalar boson
with a mass of about 38~MeV, which we called $E(38)$. The original motivation
was a decades-old geometric model\cite{Dullemond84} of quark-gluon confinement
based on an anti-De-Sitter metric, in which a light scalar field played a
fundamental role, but only in Ref.\cite{Beveren11} we noticed first indirect
experimental signals of such a boson. These amounted to small oscillations in
several production data, as well as a clear and unexplained asymmetry in
bottomonium decays (see Refs.\cite{Beveren11,Beveren20} for details and further
theoretical support). A much more direct indication we reported in
Ref.\cite{Beveren12a}, namely a very pronounced $\gamma\gamma$ peak in data
published\cite{COMPASS11} by the COMPASS collaboration, extracted by us from
those data and also shown here in Fig.~5. However, this interpretation of the
data was
\begin{figure}[!t]
\label{e38-Compass}
\begin{center}
\includegraphics[trim = 0mm 0mm 0mm 0mm,clip,width=8cm,angle=0]
{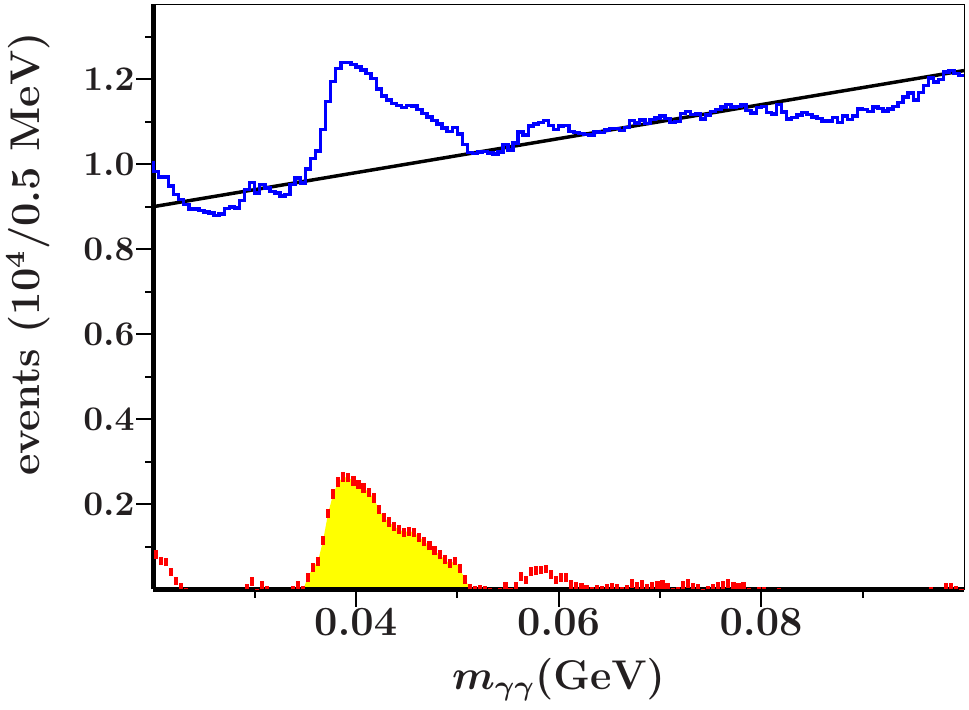}
\end{center}
\caption{Top: signal in $\gamma\gamma$ data,\protect\cite{COMPASS11}
with maximum around 39 MeV. Bottom: the structure remaining after
background subtraction and containing about 46000 events.}
\end{figure}
contested\cite{COMPASS12} by COMPASS, followed again by our
rebuttal\cite{Beveren12b} of their arguments. Nevertheless, more recently an
experimental group at the Joint Institute for Nuclear Research in Dubna 
observed\cite{Dubna19} direct $\gamma\gamma$ signals at 38~MeV in proton and
deuteron scattering off carbon and copper nuclei, though with insufficient
statistics to be considered an indisputable observation.
\section{Conclusions and outlook}
\label{concl}
As in previous published work, we have shown in this paper that the
light scalar mesons can be modelled as $q\bar{q}$ states without exotic
configurations, provided that their coupling to $S$-wave MM channels is
accounted for in a unitary framework. Thus, they emerge as a nonet of
dynamical resonances, which nevertheless owe their existence to bare
$P$-wave $q\bar{q}$ states above 1~GeV. The latter ones also become
resonances upon unitarisation, either as shifted intrinsic states or
again as dynamical ones. As for $f_0(980)$ and $a_0(980)$, they
end up as relatively narrow resonances close to and strongly attracted
by the $K\bar{K}$ threshold, also because their lowest decay modes are
suppressed. In the $f_0(980)$ case, this happens because of the mostly
$s\bar{s}$ quark content of this state, which component does not couple
to $\pi\pi$. For $a_0(980)$, the coupling to $\pi\eta$ is reduced because
of the large pseudoscalar mixing angle and is further suppressed by the
Adler zero\cite{Beveren06} just below that threshold.

In future work on the scalar mesons, we plan to generalise
Eqs.~(\ref{inter}--\ref{omega}) so as to allow for different decay radii
in all situations and will also consider alternative vertex functions.
Moreover, the empirical procedure to deal with complex asymptotic masses
in the $S$-matrix while preserving unitarity, as developed in
Ref.\cite{Coito11}, will be employed to better deal with multiparticle
decays. This may improve the phase-shift behaviour above 1~GeV.

To conclude, we stress the enormous importance of experimentally
confirming the $E(38)$ boson, which could have far-reaching consequences
for QCD and would also be a further tribute to Gribov's work on
confinement.


\begin{thebibliography}{99}
\bibitem{Gribov91}
V.~Gribov,
{\it Eur.\ Phys.\ J.} \/{\bf C10}, 71 (1999)
[arXiv:hep-ph/9807224];
91 (1999)
[arXiv:hep-ph/9902279].

\bibitem{PDG92} 
K.~Hikasa {\it et al.} [Particle Data Group],
{\it Phys.\ Rev.} \/{\bf D45}, S1 (1992)
[Erratum: {\it Phys.\ Rev.} \/{\bf D46}, 5210 (1992)].

\bibitem{Gribov93}
F.~E.~Close, Y.~L.~Dokshitzer, V.~N.~Gribov, V.~A.~Khoze, and M.~G.~Ryskin,
{\it Phys.\ Lett.} \/{\bf B319}, 291 (1993).

\bibitem{Rupp18} 
G.~Rupp and E.~van Beveren
{\it Acta Phys.\ Polon.\ B Proc.\ Supp.} \/{\bf 11}, 455 (2018)
[arXiv:1806.00364 [hep-ph]].

\bibitem{Jaffe77}
R.~L.~Jaffe,
{\it Phys.\ Rev.} \/{\bf D15}, 267 (1977); 281 (1977).

\bibitem{PDG20}
P.~A.~Zyla {\it et al.} [Particle Data Group],
{\it Prog.\ Theor.\ Exp.\ Phys.} \/{\bf 2020}, 083C01 (2020).

\bibitem{Beveren86}
E.~van Beveren, T.~A.~Rijken, K.~Metzger, C.~Dullemond, G.~Rupp,
and J.~E.~Ribeiro,
{\it Z.\ Phys.} \/{\bf C30}, 615 (1986)
[arXiv:0710.4067 [hep-ph]].

\bibitem{Zhou20} 
Z.~Y.~Zhou and Z.~Xiao,
arXiv:2008.02684 [hep-ph],
arXiv:2008.08002 [hep-ph].

\bibitem{Rupp09}
G.~Rupp, S.~Coito, and E.~van Beveren,
{\it Acta Phys.\ Polon.\ B Proc.\ Supp.} \/{\bf 2}, 437 (2009)
[arXiv:0905.3308 [hep-ph]].

\bibitem{Beveren03a}
E.~van Beveren and G.~Rupp,
{\it Int.\ J.\ Theor.\ Phys.\ Group Theor.\ Nonlin.\ Opt.} \/{\bf 11},
179 (2006)
[arXiv:hep-ph/0304105].

\bibitem{Dubna19}
K.~Abraamyan, C.~Austin, M.~Baznat, K.~Gudima, M.~Kozhin, S.~Reznikov,
and A.~Sorin,
{\it EPJ Web Conf.} \/{\bf 204}, 08004 (2019).

\bibitem{Beveren11}
E.~van Beveren and G.~Rupp,
arXiv:1102.1863 [hep-ph];

\bibitem{Beveren12a}
E.~van Beveren and G.~Rupp,
arXiv:1202.1739 [hep-ph].

\bibitem{Beveren21}
E.~van Beveren and G.~Rupp,
arXiv:2012.03693 [hep-ph], to be published.

\bibitem{Beveren80} 
E.~van Beveren, C.~Dullemond, and G.~Rupp,
{\it Phys.\ Rev.} \/{\bf D21}, 772 (1980)
[Erratum: {\it Phys.\ Rev.} \/{\bf D22}, 787 (1980)].

\bibitem{Beveren83} 
E.~van Beveren, G.~Rupp, T.~A.~Rijken, and C.~Dullemond,
{\it Phys.\ Rev.} \/{\bf D27}, 1527 (1983).

\bibitem{PDG86} 
M.~Aguilar-Benitez {\it et al.} [Particle Data Group],
{\it Phys.\ Lett.} \/{\bf B170}, 1 (1986).

\bibitem{Beveren01} 
E.~van Beveren and G.~Rupp,
{\it Eur.\ Phys.\ J.} \/{\bf C22}, 493 (2001)
[arXiv:hep-ex/0106077].

\bibitem{PDG00} 
D.~E.~Groom {\it et al.} [Particle Data Group],
{\it Eur.\ Phys.\ J.} \/{\bf C15}, 1 (2000).

\bibitem{Beveren03b} 
E.~van Beveren and G.~Rupp,
{\it Phys.\ Rev.\ Lett.} \/{\bf 91}, 012003 (2003)
[arXiv:hep-ph/0305035].

\bibitem{Beveren06} 
E.~van Beveren, D.~V.~Bugg, F.~Kleefeld, and G.~Rupp,
{\it Phys.\ Lett.} \/{\bf B641}, 265 (2006)
[arXiv:hep-ph/0606022].

\bibitem{Beveren09} 
E.~van Beveren and G.~Rupp,
{\it Annals Phys.} \/{\bf 324}, 1620 (2009)
[arXiv:0809.1149 [hep-ph]].

\bibitem{Coito09} 
S.~Coito, G.~Rupp, and E.~van Beveren,
{\it Phys.\ Rev.} {\bf D80}, 094011 (2009)
[arXiv:0909.0051 [hep-ph]].

\bibitem{Beveren82} 
E.~van Beveren,
{\it Z.\ Phys.} \/{\bf C17}, 135 (1983)
[arXiv:hep-ph/0602248].

\bibitem{Scadron13} 
M.~D.~Scadron, G.~Rupp, and R.~Delbourgo,
{\it Fortsch.\ Phys.} \/{\bf 61}, 994 (2013)
[arXiv:1309.5041 [hep-ph]].

\bibitem{Taylor72}
J.~R.~Taylor,
\em Scattering Theory: The Quantum Theory of Nonrelativistic Collisions, \em
John Wiley \& Sons, Inc., New York, London, Sydney, Toronto, 1972, pp.\ 477,
ISBN 0-471-84900-6.

\bibitem{Dudek}
J.~J.~Dudek {\it et al.} [Hadron Spectrum Collaboration],
{\it Phys.\ Rev.} \/{\bf D93}, 094506 (2016)
[arXiv:1602.05122 [hep-ph]];
R.~A.~Briceno, J.~J.~Dudek, R.~G.~Edwards, and D.~J.~Wilson,
{\it Phys.\ Rev.\ Lett.} \/{\bf 118}, 022002 (2017)
[arXiv:1607.05900 [hep-ph]];
{\it Phys.\ Rev.} {\bf D97}, 054513 (2018)
[arXiv:1708.06667 [hep-lat]];
D.~J.~Wilson, R.~A.~Briceno, J.~J.~Dudek, R.~G.~Edwards, and C.~E.~Thomas,
{\it Phys.\ Rev.\ Lett.} \/{\bf 123}, 042002 (2019)
[arXiv:1904.03188 [hep-lat]].

\bibitem{Coito11} 
S.~Coito, G.~Rupp, and E.~van Beveren,
{\it Eur.\ Phys.\ J.} \/{\bf C71}, 1762 (2011)
[arXiv:1008.5100 [hep-ph]].

\bibitem{Coito13} 
S.~Coito, G.~Rupp, and E.~van Beveren,
{\it Eur.\ Phys.\ J.} \/{\bf C73}, 2351 (2013)
[arXiv:1212.0648 [hep-ph]].

\bibitem{Dullemond84} 
C.~Dullemond, T.~A.~Rijken, and E.~van Beveren,
{\it Nuovo Cim.} \/{\bf A80}, 401 (1984).

\bibitem{COMPASS11} 
T.~Schluter [COMPASS Collaboration],
{\it eConf} \/{\bf C110613}, 83 (2011)
[arXiv:1108.6191 [hep-ex]].

\bibitem{Beveren20} 
E.~van Beveren and G.~Rupp,
arXiv:2005.08559 [hep-ph].

\bibitem{COMPASS12} 
J.~Bernhard {\it et al.} [COMPASS Collaboration],
arXiv:1204.2349 [hep-ex].

\bibitem{Beveren12b} 
E.~van Beveren and G.~Rupp,
arXiv:1204.3287 [hep-ph].

\end{thebibliography}
\end{document}